\newcommand{\slaninafiginline}[1]{}

\newcommand{\slaninafigure}[2]{
\begin{figure}[hb]
  \centering
  \vspace*{55mm}
  \includegraphics{#1.ps}
  \caption{#2}
  \label{fig:#1}
\end{figure}
}
\newcommand{\slaninafigurebig}[2]{
\begin{figure}[hb]
  \centering
  \vspace*{105mm}
  \includegraphics{#1.ps}
  \caption{#2}
  \label{fig:#1}
\end{figure}
}
\documentstyle[prl,aps,epsfig]{revtex}
\begin{document}
\draft

\twocolumn[\hsize\textwidth\columnwidth\hsize\csname@twocolumnfalse\endcsname

\title{Social organization in the Minority Game model 
}

\author{Franti\v{s}ek Slanina}
\address{        Institute of Physics,
	Academy of Sciences of the Czech Republic,\\
	Na~Slovance~2, CZ-18221~Praha,
	Czech Republic\\
        e-mail: slanina@fzu.cz
}
\maketitle
\begin{abstract}
 We study the role of imitation within the Minority Game model of
market. The players can exchange information locally, which leads to
formation of groups which act as if they were single players.
Coherent spatial areas of rich and poor agents result.
We found that the global effectivity is optimized at certain value of
the imitation probability, which decreases with increasing memory length.
The social tensions are suppressed for large imitation probability,
but generally the requirements of high global effectivity and low
social tensions are in conflict.

 \end{abstract}

\twocolumn]

\section{Introduction}
The last decade witnessed the increasing effort of physicists to formulate
various aspects of social sciences in the language of natural sciences, namely
bring them close to physics. Among numerous applications we can mention
for example traffic flow \cite{na_sch_92,na_pa_95}, city growth
\cite{ma_zha_98a}, human trail 
system \cite{he_ke_mo_97}, geometry of social networks
\cite{wa_stro_98,bar_alb_99} economic returns of research
\cite{so_zaj_99} and so  
on.

Perhaps the most prominent example of this effort is the field of 
econophysics \cite{ma_sta_99,bou_pot_00}. Among various attempts to grasp the
nature of economic behavior the modeling of systems of heterogeneous
adaptive agents reached high popularity. One of the basic models is
the Minority Game \cite{cha_zha_97,cha_zha_98} introduced by Challet
and Zhang. 

In the Minority Game (MG) we have $N$ players who choose repeatedly
between two options and compete to be in the minority 
group. This is the idealization of various situations, where the
competition for limited resources leads to intrinsic frustration. One
can think for example of cars choosing between two alternative routes
or a speculator who tries to earn money by buying and selling shares
in such a manner that 
the majority takes the opposite action than herself.

The players share a public information, saying what were the outcomes
of the game in past $M$ rounds. The players interact only through this
information. Therefore, the system has a ``mean-field'' character, in
the sense that no short-range interactions exist. 

The self-organization is achieved by allowing players to have several
strategies and choose among them the strategy which seems to be the best
one. This feature leads to decrease of the fluctuations of
attendance below its random coin-tossing value, thus increasing the
global effectivity of the system. It was found that the relevant
parameter is $\alpha=2^M/N$ and the maximum effectivity is reached for
$\alpha=\alpha_c\simeq 0.34$ 
\cite{cha_zha_98,sa_ma_ri_99,jo_ja_jo_che_kwo_hui_98}.

 It was established that a phase
transition occurs at this value of the parameter $\alpha$ and its
properties were thoroughly studied both numerically \cite{cha_ma_99} 
and analytically \cite{cha_ma_ze_00}. The Minority Game and
its variants were intensively studied by several groups. We are able
to give only an incomplete list of references
\cite{ca_pla_gu_98,ca_ga_gia_she_99,cavagna_99,jo_hu_zhe_ha_99,cha_mar_00,jo_hu_jo_lo_98,jo_hu_zhe_tai_99,ceva_99,dhu_rod_00,jo_ha_hu_98,ha_jef_joh_hu_00}.
Among the various ramifications, we would like to mention especially
the attempts to go back to the economic motivations of MG and model
the market mechanisms 
\cite{cha_ma_zha_99,sla_zha_99,jo_ha_hu_zhe_99}.

The purpose of the present work is to investigate the properties of
social structures, which can emerge if we go beyond the mean-field
character of the usual MG and allow a local information
exchange. Related works were already done, either assuming that the global
information is fully replaced by a local one \cite{kal_schu_bri_00}
or using the MG scheme for evolving the Kauffmans's boolean networks
\cite{kauffman_90a} to the critical state \cite{pa_bas_99}.

Here we want to study the effect of imitations. Indeed, it is quite
common that people do not invest individually, but rely on an advice
from specialized agencies, or simply follow the trend they perceive in
their information neighborhood. In so doing, the individuals coalesce
into groups, which act as single players. In the framework of 
Minority Game, we will study the social structure induced by the
occurence of these groups.  

\section{Minority Game with imitation}
We introduce the possibility of local information exchange in our
variant of the Minority Game. To this end, we should assume some
properties of the information network underlaying the ``information
metabolism'', in analogy to the metabolic pathways in living
organisms. The study of the geometry of information networks is now a
scientific field on its own
\cite{wa_stro_98,bar_alb_99,sla_ko_00}. Within the framework of MG a
linear chain \cite{kal_schu_bri_00} and random network with 
fixed connectivity $K$ 
\cite{pa_bas_99} was
already investigeted in different contexts. 
Here we
assume the simplest geometry, namely we will place the players on a
linear chain with nearest-neighbor connections. We will simplify the network even more by suposing,
that the information flow is not symmetric and each player can obtain
information only from her left-hand neighbor. In the rest of the
article, by ``neighbor'' we always mean the left-hand neighbor. 
Each player can read the information about her neighbors' wealth.

There wil be two conditions needed for a player to imitate her
neighbor. First, the player should have internal disposition for being
an imitator. We simplify the variety of risk-aversion levels by
postulating only two types of players. The imitator type (denoted 1) will
always adopt the strategy of her neighbor, if she finds it better. The
rest consists of the leader type (denoted 0) who always uses her own
strategies. In the whole population of $N$  
players, there is the fraction $p\in [0,1]$ of the type 1 and
the rest is of the type 0.
The case $p=0$ corresponds to the usual MG.

The second condition for the player of the type 1 to actually imitate
in the current step is that her neighbor has larger accumulated wealth
than the player itself. We suppose that the played does not know what
are the strategie of her neighbor, but if she observes that the
neighbor's behavior is more profitable than her own strategy, she
relegates the decision to the neighbor and takes the same action.
The player of the type 0 will never imitate. Therefore, she will always
look only at her $S$ strategies and choose the best estimate from
them. 

The above rules are formalized as follows. We follow mostly the
notation of \cite{cha_ma_ze_00}. There are totally $N$ players. We
will take always $N$ as an odd number. Each player has $S=2$
strategies, denoted $s_j\in\{1,2\}$. 

We denote the two possible actions a player can take $+1$ and $-1$. The
winning action is $+1$ if most players took $-1$ and vice versa.
The players know  the last $M$ outcomes of the game. This information
is arranged into the $M$-bit string $\mu\in\{-1,+1\}^M$. The strategies
are tables 
attributing to each of $2^M$ possible strings $\mu$ the action
$a_{j,s_j}^\mu$ the
player $j$ takes, if she chooses the strategy $s_j$. The scores
$U_{j,s}$ of the strategies are updated according 
to the minority rule
\begin{equation}
U_{j,s}(t+1)=U_{j,s}(t)-a_{s,j}^{\mu(t)}\,{\rm sign}\, \sum_i a_i(t) 
\end{equation}
where $a_j(t)$ is the action the player $j$ takes at time $t$.

Up to this moment the algoritm was identical to the usual MG.
We introduce a new feature, the imitation, into the rule determining
what are the actions taken by the players, depending on their wealths
and on the scores of their strategies.

Each player has a label $\tilde{l}\in\{{\rm 1},{\rm 0}\}$ indicating,
whether the player is potentially an imitator ($\tilde{l}=1$) or
always a leader ($\tilde{l}=0$). The labels are random 
but quenched. At the beginning we take each of the players and
attribute her the label 1 with probability $p$ and label 0 with
probability $1-p$. 

Denote $W_j$ the wealth of the $j$-th player.
Then, if the label $\tilde{l}_j=1$
and $W_{j-1}>W_j$, then $j$ will
take the same action as $j-1$, so $a_j=a_{j-1}$. Therefore, wee need to
establish $a_{j-1}$ first and we proceed recurrently by iterating this
rule. 
On the other hand, if 
$\tilde{l}_j=0$
or $W_{j-1}\le W_j$, the action will be $a_j=a_{j,s_{\rm M}}$,
where $s_{\rm M}$ denotes the most successfull strategy at the moment, 
$U_{j,s_{\rm M}}=\max_s U_{j,s}$, as in the usual MG.

This can be written formally by introducing the variables $l_j$
describing the actual state of imitation, in analogy with the labels
$\tilde{l}_j$ describing potential state of imitation. We can write
$l_j =\tilde{l}_j\,\theta(W_{j-1}-W_j)$, with $\theta(x)=1$ for $x>0$
and 0 othervise. The actions are
\begin{equation}
a_j = l_j a_{j-1} + (1-l_j)a_{j,s_{\rm M}}\;\; .
\end{equation}

We also suppose that the imitation is not for free. The player who
imitates passes a small fraction $\varepsilon$ of its wealth increase to
the imitated player. This rule accounts for the price of
information. We used the value $\varepsilon = 0.05$. Then, we update the
wealth of players iteratively,
\begin{equation}
\Delta W_{j}(t) = (1-\varepsilon l_j)(\varepsilon l_{j+1}\,\Delta W_{j+1}(t)-a_{j}\,{\rm sign}\, \sum_i a_i(t))
\end{equation}
where $\Delta W_j(t) = W_{j}(t+1) - W_{j}(t)$.

\slaninafigure{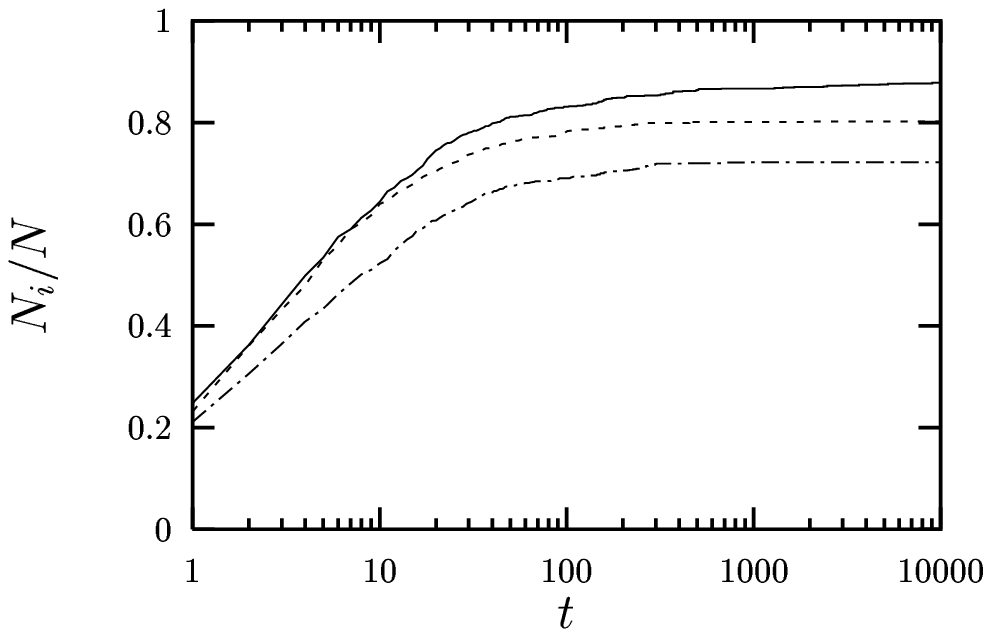}{
Time dependence of the fraction of imitating players for the number of
players is $N=1001$ and  memory $M=6$. The imitation
probabilities $p$ are 0.99 (solid line), 0.95 (dashed line), and 0.8
(dash-dotted line).}
In numerical simulations we observe that the number of actually
imitating players, $N_i=\sum_j l_j$, grows until it saturates at a
value slightly below the upper bound $p\,N$. It means that it is
always beneficiary for some potential imitators not to imitate but
rely on their own 
strategies. 
The time dependence of
the fraction of imitators $N_i/N$ for three values of $p$ is shown in
Fig. \ref{fig:imi-vs-time}.

The initially random 
distribution of wealth among players changes qualitatively during the
evolution of the system. Coherent groups of poor and wealthy players
are formed. An example of the time evolution of the spatial wealth
distribution is given in Fig. \ref{fig:structure}.

\twocolumn[
\slaninafigurebig{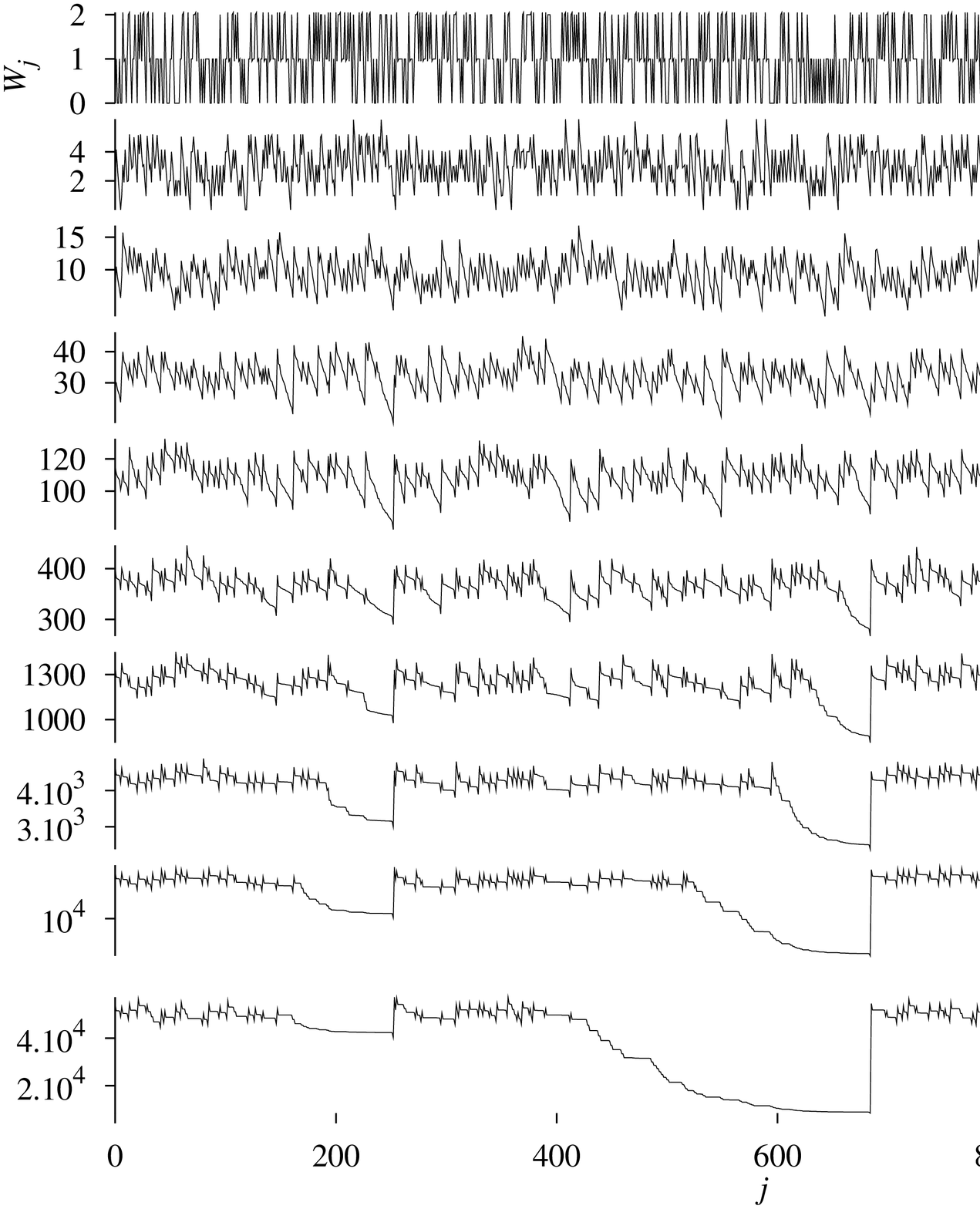}{
Example of the evolution of the distribution of wealth among
  players, for $p=1$.  The time
  step at which the 
  snapshot is taken is indicated on the right. For each time, the
  vertical axis indicates the wealth $W_j$ of the $j$'th player.}
]

\section{Global effectivity and local social tensions}

The Minority Game is a negative-sum game. The average gain of the
players reaches its maximum, if the attendance $A=\sum_j
\frac{1}{2}(a_j+1)$ is 
closest possible to $N/2$. So, the time averaged attendance fluctuations 
$\sigma^2=\langle (A-N/2)^2\rangle$ measure the distance from the
global optimum. The global effectivity is higher for smaller
$\sigma^2$. 

We investigated the influence of the imitation on the global
effectivity. In the usual MG (which corresponds to $p=0$) the fluctuations
depend on the memory and 
number of players through the scaling variable $\alpha=2^M/N$. 
This
fact can be observed through the
data collapse of the results for various $N$'s. We
tried to find the same data collapse for a non-zero $p$. Figure
\ref{fig:sig-vs-m-alln-0.8} shows such a plot for $p=0.8$. We can see that the
dependence of $\sigma^2/N$ on $\alpha$ has a minimum as in the usual MG,
but the critical value $\alpha_c$ of the scaling parameter is lower
while the value of the flucuations at the minimum is higher. In the
symetry-broken phase ($\alpha>\alpha_c$) the data collapse is good. On
the other hand, significant deviations are observed in the crowded
(symmetric) phase, which are more
pronounced for smaller $\alpha$.
On the other hand, for larger value of imitation probability, the data
still scale with $\alpha=2^M/N$, but the collapse is poorer in the
whole range of $\alpha$. The situation  
for $p=0.99$ is shown in Fig. \ref{fig:sig-vs-m-alln-0.99}. We can
see that if we decrease $\alpha$ in the symmetric phase, the
fluctuations increase first, but contrary to the usual MG a local
maximum is reached and when decreasing $\alpha$ further, the
fluctuations are suppressed. 
\slaninafigure{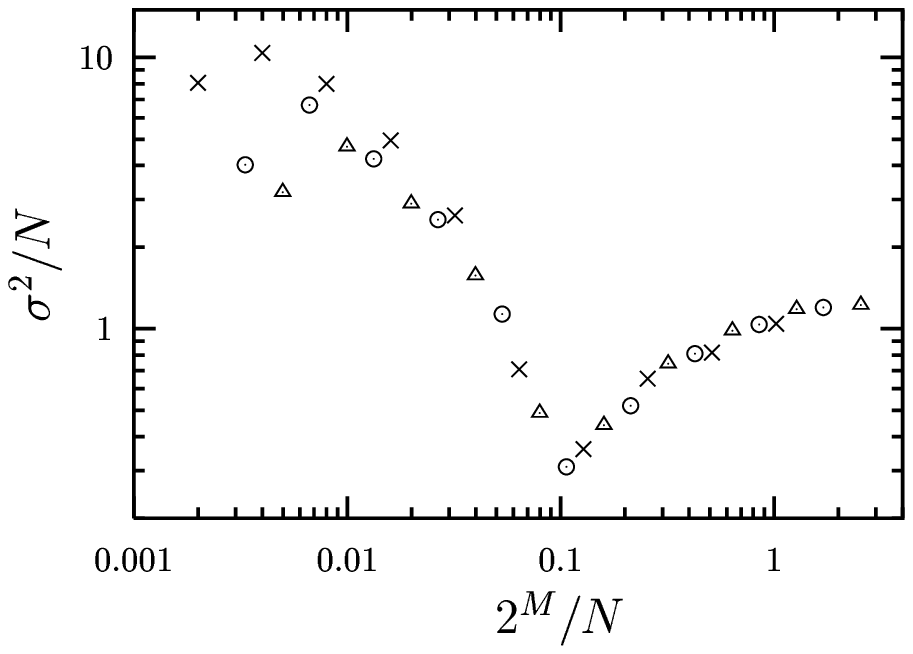}{Dependence of the attendance
 fluctuations on the scaling variable $2^M/N$, for $p=0.8$ and system
sizes $N=1001$ ($\times$), 601 ($\odot$), and 401 ($\triangle$).
}
\slaninafigure{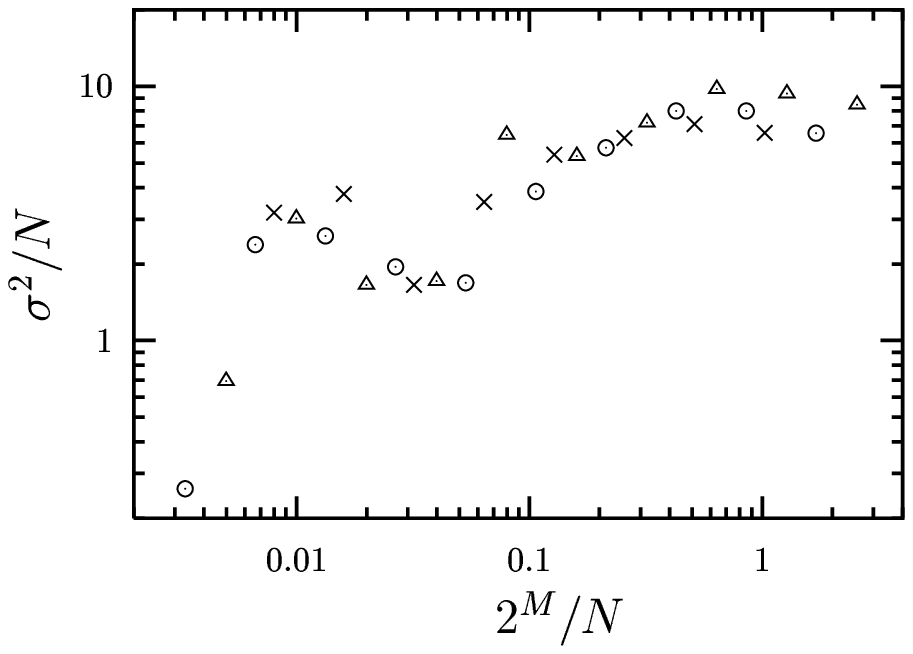}{Dependence of the attendance
 fluctuations on the scaling variable $2^M/N$, for $p=0.99$ and system
sizes $N=1001$ ($\times$), 601 ($\odot$), and 401 ($\triangle$).
}

We tried also another data collapse, namely the dependence of
$\sigma^2/Ns_\sigma$ on $2^Ms_M/N$ for various $p$, where $s_M$ and
$s_\sigma$ are scaling parameters depending on $p$, chosen so that 
the best possible
data collapse is achieved. For convenience we require $s_M=s_\sigma=1$
for $p=0$. With this choice, the scaling parameters have the following
interpretation. The location minimum of attendance fluctuations is shifted 
from its usual value $\alpha_c(0)=0.34$ valid for $p=0$ to 
$\alpha_c(p)=\alpha_c(0)/s_M$, while the value of $\sigma^2/N$ in the
minimum is increased by the factor $s_\sigma$. We can see the results
in Fig. \ref{fig:sig-vs-m-1001-allp}. We again observe good collapse
in the symmery-broken regime, while in the symmetric phase there is no
collapse, but systematic decrease, stronger for higher $p$, is observed.  
\slaninafigure{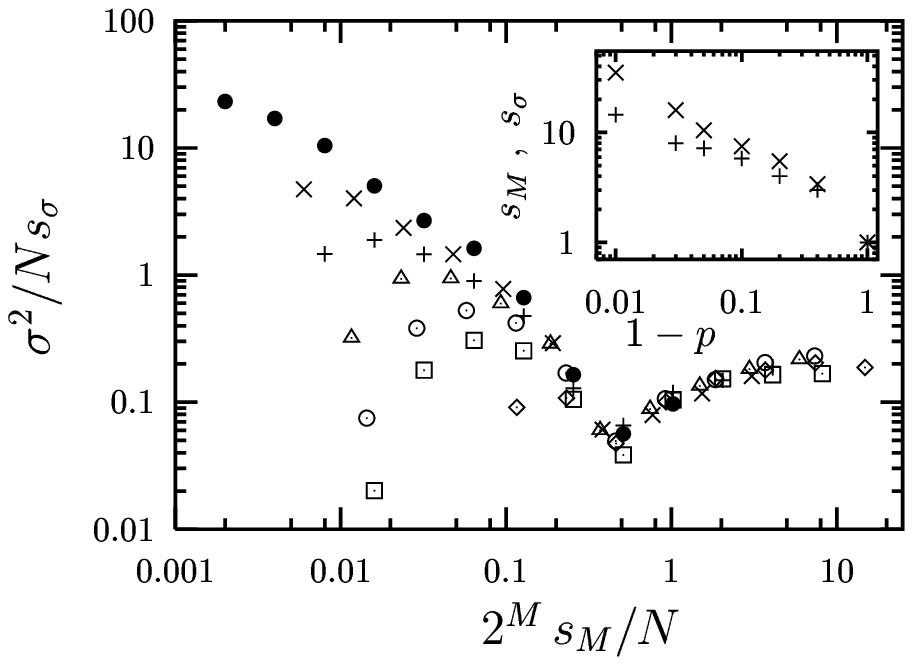}{Rescaled atendance fluctuations. The
 number of players is $N=1001$ and the imitation probabilities $p$ are 0
 ($\bullet$), 0.6 
 ($\times$), 0.8 ($+$), 0.9 ($\triangle$), 0.95 ($\odot$), 0.97
 ($\Box$), and 0.99  ($\Diamond$). Inset: dependence of  the scaling 
parameters $s_M$ (+) and $s_\sigma$ ($\times$) on $p$.}

Nevertheless, the parameters $s_M$ and $s_\sigma$ still bear the
information about the shift of the minimum. We can see their
dependence on $p$ in the inset of Fig. \ref{fig:sig-vs-m-1001-allp}.
The data suggest that both of the parameters diverge as a power law
when $p\to 1$.
 
We investigated also the dependence of the attendance fluctuations (and
therefore the global effectivity) on the imitation probability for
fixed number of players and memory. We found that in the crowded
phase the system becomes more effective if imitation is allowed
($p>0$), but there is a local minimum in the dependence of $\sigma^2/N$
on $p$, indicating that there is an optimal level of imitation,
beyond which the system starts to perform worse. The results for
$N=1001$ are shown in Fig.  
 \ref{fig:sig-all-all}. We can see that the minimum occurs at smaller
values for larger $M$. We can also observe that for longer memories
($M=7$ in our case) the value of the fluctuations for $p=1$ is
significantly above the value vithout imitation ($p=0$), while the
value at the minimum still lies below the $p=0$ value. This implies
that moderate imitation can be beneficiary, while exaggerated one can
be harmful.
\slaninafigure{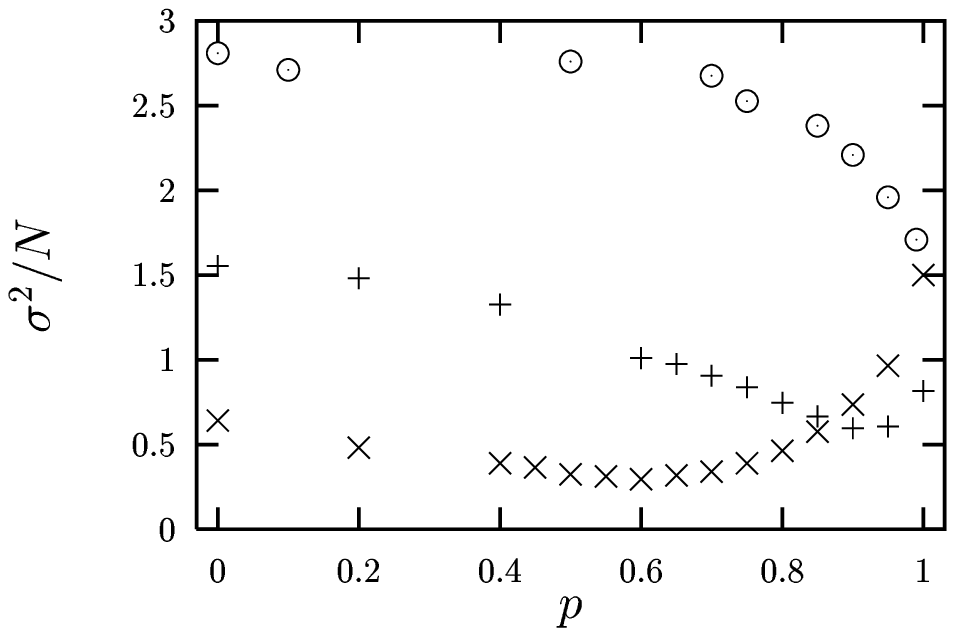}{Dependence of the attendance fluctuations
 on the imitation probability.
 The number of players is
 $N=1001$ and 
 memory length $M=5$  ($\odot$), $M=6$ ($+$), and $M=7$ ($\times$).}

The increase of spatial coherence in poor and wealthy groups seen
qualitatively in Fig. \ref{fig:structure} can result in decrease
of social tension. Indeed, the individuals can exchange information
about their wealth only among neighbors. If most of the individuals
perceive small local wealth gradient, the social tensions will be also
small. To quantify the social tension, we should introduce a sort of
``utility function'' \cite{merton_90} $U(\Delta W)$, which indicates,
how much the wealth difference $\Delta W$ is subjectively
perceived. As usual, any realistic utility function should have
negative second derivative. This feature, called risk aversion in
standard economics literature, expresses in our case the intuitively
obvious fact that smaller differences in living standards are
relatively more significant than larger differences: I could be more
happy if I succeeded in shrinking the wage distance to my rival
employee from 500~\$ to 400~\$ than from 55,500~\$ to 55,400~\$.

\pagebreak
We will use the utility function in the form
$U(x)=x^b$. Realistic
values of the exponent are $b<1$. However, we will use also the values
$b=1$ and $b=2$ for comparison. Then, the global measure of the local
social tension is
\begin{equation}
d_b=\frac{1}{\langle
W\rangle}\left(\sum_{j=1}^{N-1} |W_j-W_{j+1}|^b \right)^{1/b} 
\end{equation}
where we denoted the average wealth $\langle W\rangle=\frac{1}{N}\sum_{j=1}^{N} W_j$.

The results for various values of the exponent $b$ are shown in
Fig. \ref{fig:differences-all-all}. We can see that the behavior for
all $b$ except $b=2$ is qualitatively similar. When increasing $p$,
the social tension increases first, but then decreases again and for
$p\to 1$ it reaches a point below its value
for $p=0$. Hence, large enough imitation will decrease the social
tension. On the other hand, for $b=2$ we observe further increase 
when $p$ approaches to 1. Therefore, if the utility was measured by
exponent $b=2$ (which would hold for risk-seeking adventurers), the
tension would only increase due to imitation. 

\slaninafigure{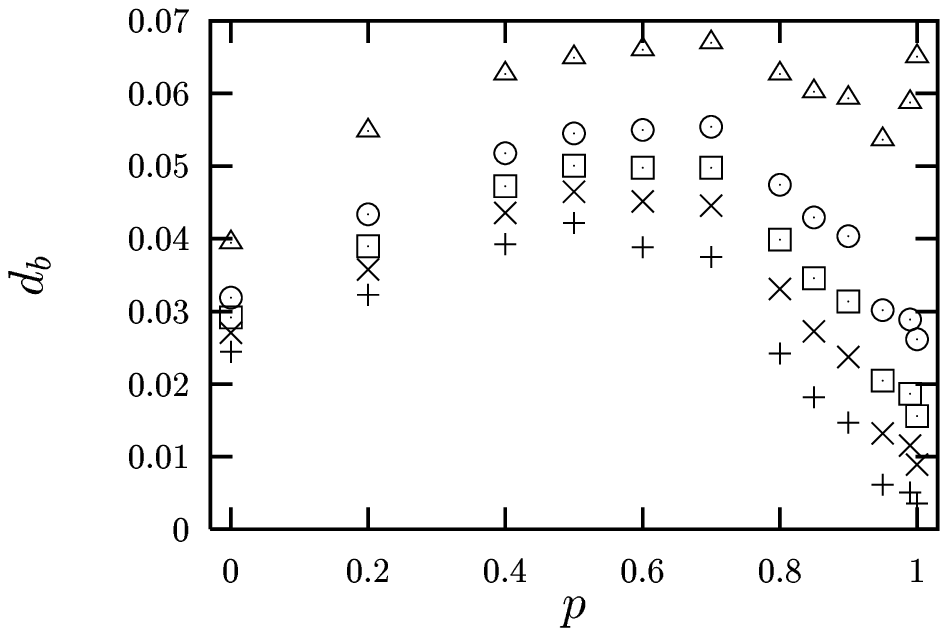}{Relative local tension for $N=1001$, $M=6$, measured
 by utility  function   $(\Delta W)^b$
 with  exponents $b=0.3$ ($+$), $b=0.5$ ($\times$), $b=0.7$ ($\Box$),
 $b=1$ ($\odot$), and 
 $b=2$ ($\triangle$).}
The most striking result comes from comparison of Fig
\ref{fig:sig-all-all} (for $M=6$) and Fig
\ref{fig:differences-all-all}. We can 
see that both the local maxima of $\sigma^2/N$ and the local minima of
$d_b$ for $b<1$ lie at the endpoints of the interval [0,1]. Therefore,
if we want to optimize the social tensions by some process invoving
small gradual changes of $p$, we will finish in one of the endpoints,
depending on the initial value of $p$. 
Among the two stable points, the optimum is located at $p=1$.

However, the stable points (including the absolute optimum of the
local tension) are simultaneously
local minima of the global effectivity: we observe a clear conflict of
interest between local and global views.   

\section{Conclusions}
We investigated the creation of rich and poor spatial domains due to
local information exchange, within the framework of the Minority
Game. The introduction of imitation among players leads to creation of
groups, which act in accord. For fixed imitation probability $p$ we
found that the attendance fluctuations still depend on the scaling
variable $\alpha=2^M/N$ in the symmery-broken phase, while in the symmetric
phase there are systematic deviations from the data collapse. 

The characteristic minimum in  the dependence of fluctuations on
$\alpha$ persist in our modification of the Minority Game, but the
location of the minimum decreases and the value at the minimum increases
when increasing  $p$. We found that
both quantities depend as a power law on $1-p$.

We found that in the symmetric phase the introduction of imitation
can lead to absolute decrease of the attendance fluctuations and thus
to increase in global system effectivity. There is an optimal value of
the imitation, which increases with decreasing $M$.

The creation of coherent areas of poor and rich agents leads to
decrease in the local social tensions, but only if $p$ is sufficiently
close to 1. The lowest value of the social tension is reached at
$p=1$, but for such a value the global effectivity is significantly
lower than its optimum value. Therefore, we observe a conflict of
local interests (minimization of social tension) with global
performance (minimization of attendance fluctuations).
\acknowledgments{I am indebted to Yi-Cheng Zhang for numerous useful discussions. I
acknowledge the financial support from the University of Fribourg,
Switzerland, where part of this work was done.  
 
}


\end{document}